\documentclass[graybox, envcountchap]{my_svmult}
\usepackage{mathptmx}        
\usepackage{amsmath}
\usepackage{amssymb}
\usepackage{color}
\usepackage{helvet}          
\usepackage{courier}         
\usepackage{dirtree}

\usepackage{makeidx}        
\usepackage{graphicx}        
\usepackage{subfig}

\usepackage{multicol}        
\usepackage[bottom]{footmisc}

\usepackage{hyperref}        
\hypersetup{colorlinks=true,urlcolor=blue}

\usepackage[misc]{ifsym}

\usepackage{sidecap}
\usepackage{natbib}
\usepackage{tikz}
\usetikzlibrary{arrows.meta}

\makeindex             

\begin{document}


\title{Machine Learning Techniques for Astrophysics and Cosmology: Simulation-Based Inference}
\titlerunning{Simulation-Based Inference}
\author{Leander Thiele}
\institute{Leander Thiele (\Letter) \at Kavli IPMU (WPI), UTIAS, The University of Tokyo, 5-1-5 Kashiwanoha, Kashiwa, Chiba 277-8583, Japan, \email{leander.thiele@ipmu.jp}}
%
%
\maketitle

\abstract{
  Simulation-based inference (SBI) enables parameter inference by training neural networks on forward simulations.
  It is being applied both for intractable likelihoods as well as under time constraints on the posterior sampling.
  After motivating situations in which SBI is useful, we give a pedagogical description of the basic techniques.
  These are posterior, likelihood, and ratio estimation.
  Alternatives, sequential versions, and learned summaries are discussed briefly.
  We provide a brief guide to choosing among the techniques in practical scenarios.
  SBI needs to be verified through diagnostics since failures can be subtle but would invalidate the inference result.
  We explain the most common diagnostic techniques.
  We briefly list some recent SBI applications in the cosmology and astrophysics literature.
  Before concluding, we discuss current methodological challenges.
  We identify training with limited simulation budgets as the critical problem for applications to cosmology and astrophysics.
}


\section{Motivation}

Simulation-based inference\index{simulation-based inference}\index{SBI} (SBI) is a collection of techniques to estimate model parameters using a training set consisting of forward simulations of the model.
In the Bayesian paradigm, parameter estimation amounts to computation of the \emph{posterior}\index{posterior}
\begin{equation}
  p(\theta | x_o) = \frac{p(x_o | \theta) p(\theta)}{p(x_o)}\,,
\end{equation}
where $\theta$ is the parameter vector to estimate, $x_o$ is the observed data vector, $p(\theta)$ is the prior\index{prior}, $p(x)$ the evidence, and $p(x | \theta)$ the likelihood\index{likelihood}.
While the evidence only controls normalization independently of $\theta$, and the prior is decided on by the analyst, the likelihood encapsulates the physical process connecting model parameters with observable phenomena.
There are two motivations for using SBI to estimate the posterior, both having to do with the likelihood.

\begin{figure}
  \centering
  \begin{tikzpicture}[
    font=\sffamily,
    >=Stealth,
    line/.style={draw, -{Stealth[length=3mm]}, thick},
    txt/.style={align=left}
  ]
  \draw[line] (-2.5,0) -- (2.5,0);
  \draw[line] (-1.5,0.7) -- (-1.5,0);
  \draw[line] (1.5,0.7) -- (1.5,0);
  \draw[line]
    (2.5,-0.3) to[out=-160,in=-20] 
    node[txt, below] {inference of posterior $p(\theta \mid x)$}
    (-2.5,-0.3);
  \node[txt, anchor=east] at (-2.5,0)
    {parameters $\theta \sim p(\theta)$,\\
     e.g.\ $S_8$, $H_0$, dark energy};
  \node[txt, anchor=south east] at (-0.7,0.7)
    {initial conditions $\zeta \sim p(\zeta)$,\\
     e.g.\ quantum fluctuations};
  \node[txt, anchor=south west] at (0.7,0.7)
    {nuisance parameters $\eta \sim p(\eta)$,\\
     e.g.\ star formation model};
  \node[anchor=north] at (0,-0.05)
    {physical process $m(\theta,\eta,\zeta)$};
  \node[txt, anchor=west] at (2.5,0)
    {observable data $x$,\\
     e.g.\ catalog of galaxies};
  \end{tikzpicture}%
  \caption{Schematic illustration of the setup.}
  \label{fig:schema}
\end{figure}
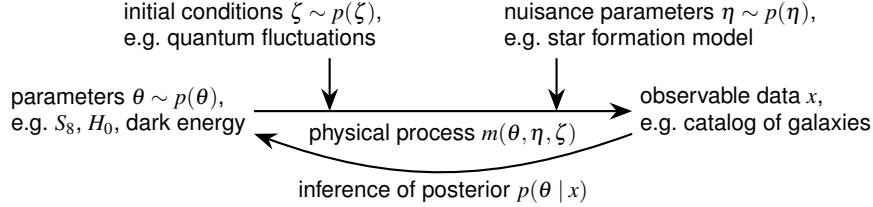

In the first situation, the likelihood itself is intractable.
Let us contrast this with more traditional explicit-likelihood inference.
Schematically, we can write the likelihood as
\begin{equation}
  p(x | \theta) = \int D\eta\,D\zeta\,\delta[x - m(\theta, \eta, \zeta)]\,,
  \label{eq:like}
\end{equation}
where $m$ is the forward process which depends on parameters $\theta$, nuisance parameters $\eta$, and initial conditions $\zeta$ (see Fig.~\ref{fig:schema}),
and the notation $D$ indicates that the measures include priors for compactness.
Initial conditions, in this language, are meant to also include experimental noise and other high-dimensional quantities impacting the measurement.
The schematic Eq.~\eqref{eq:like} simply counts which combinations of model inputs give rise to the observed data vector.
Thankfully, in practice we rarely have to work with this complicated summation directly.
Instead, it is often the case that the integral over the high-dimensional $\zeta$ can be performed exactly or to a very good approximation as
\begin{equation}
  p(x | \theta) = \int D\eta\,\text{Gaussian}[x-\mu(\theta, \eta), \Sigma]\,,
  \label{eq:gaussian}
\end{equation}
where $\mu(\theta, \eta) = \langle m(\theta, \eta, \zeta) \rangle_\zeta$ is the expected data vector and $\Sigma$ is a covariance matrix which can be computed either analytically or from simulations.
In some cases the integration over $\zeta$ can be performed to yield other integrands, for example Poisson.
If the simplified Eq.~\eqref{eq:gaussian} is valid, traditional explicit-likelihood inference techniques apply.
As $\eta$ is low-dimensional (at least compared to $\zeta$), and computer codes exist for $\mu(\theta, \eta)$, the remaining integral can be well approximated using Markov Chain Monte Carlo\index{Markov Chain Monte Carlo}\index{MCMC} (MCMC) sampling.
However, the simplification to Eq.~\eqref{eq:gaussian} is not generally accurate.
In such cases, we have to confront an intractable likelihood.
Let us assume that we do have access to a simulator which implements the forward model $m(\theta, \eta, \zeta)$ with sufficient accuracy.
Instead of going through the daunting task of trying to perform the extremely high-dimensional integral Eq.~\eqref{eq:like} directly, we can use neural networks\index{neural network} to approximate quantities involving the likelihood.

In the second situation motivating the use of SBI we are not confronted with an unknown likelihood but rather with an expensive one.
That is, the simplified Eq.~\eqref{eq:gaussian} is valid, but actually performing the marginalization over $\eta$ and posterior sampling is too expensive.
Usually, this problem arises whenever posteriors need to be estimated many times (such as in astrophysics) or in real time (such as in particle physics).
It turns out that also in this situation it can be advantageous to rely on neural networks to provide posterior estimates.

\section{Techniques}

SBI is a broad field. For the purposes of this review we focus on the three main techniques, namely neural posterior, likelihood, and ratio estimation.
Alternatives, as well as the sequential variants, will be discussed briefly.

For the following discussion, we assume that we have prepared a training dataset $\mathcal{D}$ which consists of pairs $(\theta, x)$ of parameter and data vectors which are sampled from the joint $p(\theta, x)$.
In practice, this usually means to first sample $\theta \sim p(\theta)$ (as well as nuisance parameters $\eta$ and initial conditions $\zeta$) from the prior and then run the forward simulator $m(\theta, \eta, \zeta)$ on them.
Note that it is \emph{not necessary} to draw independently and identically distributed (i.i.d.) from the prior when constructing the training set.
As we will see below, all SBI methods rely on approximating loss functions of the integral form
\begin{equation*}
  \int d\theta\,dx\,p(\theta, x) [\cdots] \approx \frac{1}{|\mathcal{D}|} \sum_{\theta, x \in \mathcal{D}} [\cdots]\,,
\end{equation*}
as sums over the training set (usually in mini-batch form).
For these sums to be unbiased estimators of the desired integral, it is not required for the samples to be i.i.d.
In fact, integral estimators with lower variance can be constructed through quasi-random sequences (such as Latin hypercube or Sobol).
The practical relevance of this point lies in the hierarchical nature of many simulators.
For example, in cosmology we might have a gravity-only simulation described by cosmological parameters, into which galaxies are inserted as described by galaxy-halo connection parameters.
One can run many realizations of the galaxy model for each gravity-only simulation and still obtain unbiased loss function estimators.

In the rest of this chapter, we use a simple but non-trivial example problem for illustration (see Fig.~\ref{fig:npe} etc.).
Both the data vector $x$ and the parameter vector $\theta$ are one-dimensional, and their joint density follows the two-moons distribution.
The prior $p(\theta)$ is non-trivial and shown in the bottom of the plots.
We choose $|\mathcal{D}|=256$ training samples which are shown as points.
All neural networks are trained as implemented in the \texttt{sbi} package~\cite{tejero-cantero2020sbi}.
We do not perform any hyperparameter optimization, and thus with the small training set the agreement between inferred posteriors and the ground truth is only fair.

\subsection{Neural Posterior Estimation}\label{sec:npe}

\begin{SCfigure}
  \includegraphics[width=0.7\textwidth]{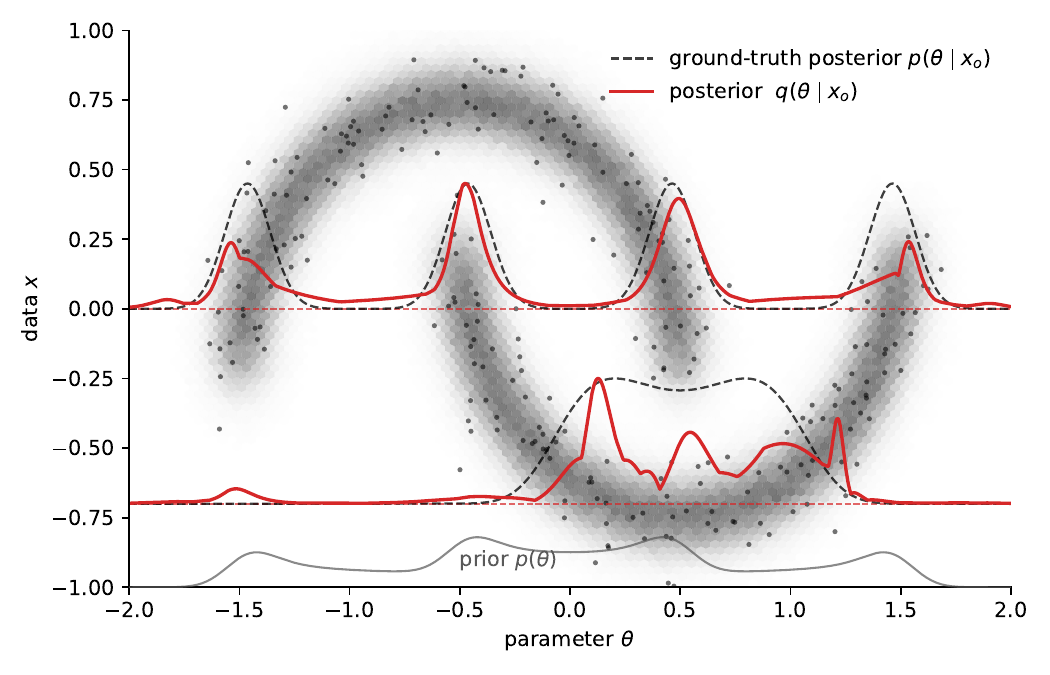}
  \caption{Illustration of posterior estimation. The normalizing flow directly predicts the solid red posteriors.}
  \label{fig:npe}
\end{SCfigure}

Neural posterior estimation\index{neural posterior estimation}\index{NPE} (NPE) targets the posterior $p(\theta|x)$ directly~\cite{Greenberg2019,Lueckmann2017,Papamakarios2016}.
In practice, we introduce a surrogate density $q_\phi(\theta|x)$ which is parameterized by a neural network with parameters $\phi$ (in many fields such a surrogate is known as a variational ansatz).

This density can be conveniently expressed through a discrete-time normalizing flow\index{normalizing flow}~\cite{Tabak:2010vsy,Tabak:2013cnz,JimenezRezende2015,Papamakarios2021}.
Such traditional normalizing flows consist of a series of invertible transformations with tractable Jacobians which transform a sample $\theta$ from the complicated target density to a sample $\theta'$ which is drawn from a simple base density (often Gaussian).
Conditioning on $x$ is done for each of the transformations.
The challenge in constructing discrete-time normalizing flows is to attain sufficient expressivity.
Continuous-time flows\index{flow matching}\index{stochastic interpolant}~\cite{Lipman2023,Liu2022,Albergo2023,Chen2024,Wildberger2023} have thus become the standard for many applications.
In the context of simulation-based inference, though, the generative problem is usually not particularly complex -- posteriors in cosmological and astrophysical applications are typically quite simple.
Given the convenience of direct amortized access to the density, discrete-time flows continue to be widely used for NPE.

Optimization of the surrogate density is performed according to the objective
\begin{equation}
  \mathcal{L}[\phi] = \langle - \log q_\phi(\theta|x) \rangle_{\theta,x \in \mathcal{D}}\,.
\end{equation}
In the limit of an infinite training set, this empirical objective approaches
\begin{equation}
  \bar{\mathcal{L}}[\phi] = - \int dx\,d\theta\,p(x) p(\theta|x) \log q_\phi(\theta|x)\,.
\end{equation}
Taking $q_\phi$ as infinitely expressive, we can find its value at optimum by functionally differentiating with respect to it.
We impose the normalization condition using a Lagrange multiplier.
Since $x$ is only a conditioning variable, we can optimize separately for each $x$ and thus obtain the objective
\begin{equation*}
  \bar{\mathcal{L}}_x[q_\phi] = - \int d\theta\,p(\theta|x) \log q_\phi(\theta|x) + \lambda_x \int d\theta\,q_\phi(\theta|x)\,,
\end{equation*}
from which we compute the stationary point as
\begin{equation*}
  \frac{\delta\bar{\mathcal{L}}_x[q_\phi]}{\delta q_\phi} = - \frac{p(\theta|x)}{q_\phi(\theta|x)} + \lambda_x = 0\,.
\end{equation*}
From the normalization condition we obtain $\lambda_x=1$ and thus the required result
\begin{equation}
  q_\phi(\theta|x) = p(\theta|x)
\end{equation}
at optimum.

A useful alternative calculation method decomposes the objective
\begin{align}
  \bar{\mathcal{L}}[\phi] &= - \int dx\,d\theta\,p(x) p(\theta|x) \log q_\phi(\theta|x) \nonumber \\
  &= \int dx\,d\theta\,p(x) p(\theta|x) \log \frac{p(\theta|x)}{q_\phi(\theta|x)} - \int dx\,d\theta\,p(x) p(\theta|x) \log p(\theta|x) \nonumber \\
  &= \langle D_\text{KL}[p(\theta|x), q_\phi(\theta|x)] \rangle_{x\sim p(x)} + \text{const}
\end{align}
into the average Kullback-Leibler divergence\index{Kullback-Leibler divergence} between target distribution and variational ansatz, and a constant term which is just the expected Shannon entropy of the posterior.

An example application of NPE to the two-moons problem is illustrated in Fig.~\ref{fig:npe}.
Example posteriors, for the observations $x_o$ indicated by dashed red lines, are shown in the solid red curves and compared against ground-truth posteriors in dashed black.

\subsection{Neural Likelihood Estimation}\label{sec:nle}

\begin{SCfigure}
  \includegraphics[width=0.7\textwidth]{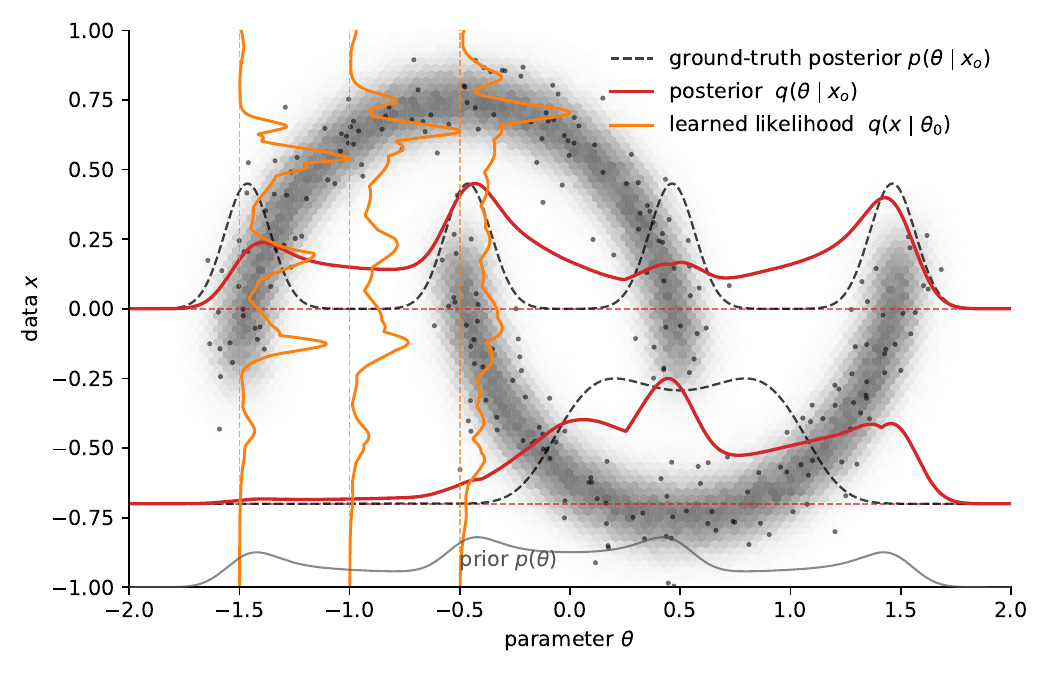}
  \caption{Illustration of likelihood estimation. The normalizing flow predicts the orange curves from which the posterior follows after MCMC sampling.}
  \label{fig:nle}
\end{SCfigure}

Neural likelihood estimation\index{neural likelihood estimation}\index{NLE} (NLE) targets the likelihood $p(x|\theta)$~\cite{Papamakarios2019nle,Lueckmann2019nle}.
The posterior can by approximated by taking the likelihood's product with the prior and MCMC sampling it.
Since the likellihood $p(x|\theta)$ is another conditional density, techniques carry over from NPE and we refer the reader to the previous Sec.~\ref{sec:npe} for a discussion of architectures.

The optimization objective in NLE is the same as in NPE, namely
\begin{equation}
  \mathcal{L}[\phi] = \langle - \log q_\phi(x|\theta) \rangle_{\theta,x \in \mathcal{D}}\,.
\end{equation}
Similar techniques as in the previous Sec.~\ref{sec:npe} can be used to show that in the limit, the optimal $q_\phi$ equals the target likelihood.

The application of NLE to the previously introduced two-moons example is shown in Fig.~\ref{fig:nle}.
Note how the direct outputs of NLE are the densities shown in vertical orange.
Thus, in contrast to NPE, MCMC sampling is needed to obtain the posterior samples.

\subsection{Neural Ratio Estimation}

\begin{SCfigure}
  \includegraphics[width=0.7\textwidth]{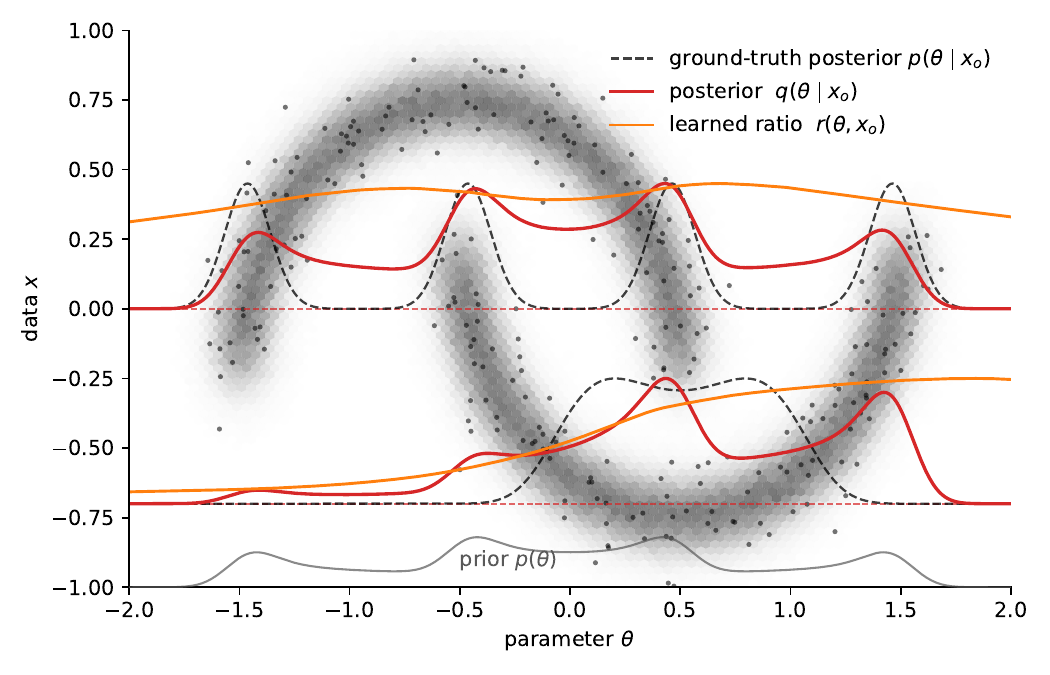}
  \caption{Illustration of ratio estimation. The classifier predicts the orange curves from which the posterior follows after MCMC sampling.}
  \label{fig:nre}
\end{SCfigure}

Neural ratio estimation\index{neural ratio estimation}\index{NRE} (NRE) targets the likelihood-to-evidence ratio $r(\theta, x) \equiv p(x|\theta)/p(x)$~\cite{Hermans2020nre,Durkan2020}.
Like NLE, the posterior can be approximated by taking the ratio's product with the prior and MCMC sampling it.
In contrast to NPE and NLE, the ratio $r(\theta, x)$ can be modeled with a classifier\index{classifier}.
Classifiers are much more flexible and usually more efficient architectures than the discrete-time normalizing flows often used for NPE/NLE as they only need to compute a forward transformation.

The classifier $f_\phi$ takes as input a concatenation of parameters and data and outputs a single scalar indicating the class.
The classifier's task is to differentiate between inputs $\theta, x$ coming from two different distributions:
either from the joint $p(\theta, x) = p(x|\theta)p(\theta)$, or from the product of marginals $p(\theta)p(x)$.
In practice, inputs from the latter distribution can be constructed by randomly pairing data and parameters across the training set.
Usually binary cross entropy is used as the classifier's objective, so that in the infinite data limit we obtain
\begin{equation}
  \bar{\mathcal{L}}[\phi] = \int dx\,d\theta\,p(\theta) [- p(x|\theta) \log f_\phi(\theta, x) - p(x) \log(1-f_\phi(\theta, x))]\,.
\end{equation}
Assuming again infinite expressivity of $f_\phi$, we compute the stationary point as
\begin{equation*}
  \frac{\delta\bar{\mathcal{L}}_x[f_\phi]}{\delta f_\phi} = p(\theta)\left[- \frac{p(x|\theta)}{f_\phi(\theta, x)} + \frac{p(x)}{1-f_\phi(\theta, x)}\right] = 0\,,
\end{equation*}
which yields at optimum the required result for the likelihood-to-evidence ratio
\begin{equation}
  r(\theta, x) \equiv \frac{p(x|\theta)}{p(x)} = \frac{f_\phi(\theta, x)}{1 - f_\phi(\theta, x)}\,.
\end{equation}
In practice, usually the logarithm of this ratio is targeted directly as the neural network's output.

The application of NRE to the previously introduced two-moons example is shown in Fig.~\ref{fig:nre}.
Note that NRE requires MCMC sampling to obtain the posterior samples, similar to NLE.

\subsection{Alternatives}

Simulation-based inference has a long history and was not always synonymous with neural network methods.
Early works developed approximate Bayesian computation (ABC)\index{approximate Bayesian computation}\index{ABC}~\cite{Pritchard1999-sa,Marjoram2003,Marin2011,Sisson2018,Grazian2019}.
In this framework, posteriors are directly approximated by histogramming simulation samples within a neighborhood of the data vector.
We can understand this technique as the low-bias, high-variance end of the spectrum (assuming the neighborhood is chosen sufficiently small).
In contrast, neural methods exploit inductive biases about regularity in order to decrease variation in the posterior.
In high dimensions and with sufficiently large training sets, we may expect that feature learning takes place and further helps with performance.

In terms of neural methods, in addition to posterior, likelihood, and ratio estimation as explained above, it has recently been proposed to target quantiles~\cite{Jia2024a}.
Such neural quantile estimation\index{neural quantile estimation}\index{NQE} (NQE) shares with NRE the distinction that it does not rely on normalizing flows and can thus use more expressive architectures.
Furthermore, it has been argued that NQE simplifies the posterior calibration (see Sec.~\ref{sec:coverage})~\cite{Jia2024b}.

\subsection{Sequential techniques}\label{sec:sequential}

In our descriptions of NPE, NLE, and NRE, we have only considered the amortized setup.
In this scenario, a definite set of simulations sampling a fixed prior has been run and is then fixed.
The setup is ``amortized'' in the sense that once the neural network is trained on this fixed simulation set, it can be used to compute posteriors for any data vector.

In contrast, sequential techniques use a specific data vector (the observation) to inform the simulation procedure.
Roughly speaking, an approximation to the posterior is computed using a first set of simulations.
Then, this first approximation is used to narrow the region where a second round of simulations is run.
A new posterior approximation is found, informs a third round of simulations, and so on.

Sequential techniques shine in situations where the posterior is much more informative than the prior.
They suffer from two problems: Before narrowing down the simulation prior, one must be very sure not to miss any important probability mass in the approximate posterior.
Furthermore, sequential inference makes it impractical to run the empirical calibration tests (see Sec.~\ref{sec:coverage}).

Thus, whether a sequential method should be preferred over an amortized one is highly dependent on the specific problem.
In cosmology, we believe that there are many problems in which the added information from simulation-based inference, as compared to previous explicit-likelihood analyses, is rather of order unity than very large.
In such situations it seems to be more reasonable to retain the additional consistency checks that amortized methods allow.

\subsection{(Neural) summaries}

In many applications, data vectors (and sometimes parameter vectors) are very high dimensional.
Since learning in high dimensions requires a large number of simulations which may not always be available, compression is often required.

General techniques for data compression\index{data compression} can be used prior to training the inference network, regardless of the choice of SBI framework (NPE, NLE, NRE, \ldots).
Linear techniques such as MOPED~\cite{Heavens2000} or the more general score compression~\cite{Alsing2018} and information-maximizing neural networks~\cite{Charnock2018} are often successful in retaining a large fraction of the available information.

To find the truly optimal compression, however, it is necessary to jointly train the compressor with the inference network.
In such cases, the compressor is known as an \emph{embedding network}\index{embedding network}.
In the usual setup in which the data vector is being compressed, joint training is not possible in NLE (since the NLE flow is in data space and thus needs to remain bijective).
Conversely, in the (rare) situation in which the parameter vector is being compressed, joint training is not possible in NPE.
As a classifier architecture, NRE can deal with both situations.

In addition to high dimensionality, data vectors often have special properties which motivate constraints on the embedding networks.
For example, cosmological field-level analyses should take into account the symmetry and regularity properties of the data, often using a convolutional network\index{convolutional neural network}\index{CNN}~\citep[e.g.,][]{Lemos2024}.
Another example is inference from object catalogs whose permutation invariance should be respected~\citep[e.g.,][]{Wang2025}.

\subsection{Practical guidelines}

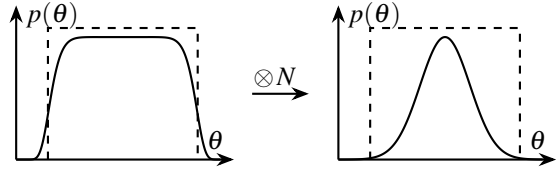
\begin{SCfigure}
    \begin{tikzpicture}[
    >=Stealth,
    thick,
    axis/.style={->, thick},
    curve/.style={thick},
    dashedbox/.style={thick, dashed},
    xscale=0.6,yscale=0.6
  ]
  \begin{scope}[shift={(0,0)}]
    \draw[axis] (0,0) -- (4.8,0) node[above left] {$\theta$};
    \draw[axis] (0,0) -- (0,3.4) node[below right, yshift=+5pt] {$p(\theta)$};
    \draw[dashedbox] (0.7,0) -- (0.7,2.9) -- (4.0,2.9) -- (4.0,0);
    \draw[curve, domain=0:4.5, samples=250, smooth]
      plot (\x,{2.7*exp(-((\x-2.35)/1.65)^10)});
  \end{scope}
  \node at (5.7,1.8) {$\otimes\, N$};
  \draw[->, thick] (5.2,1.45) -- (6.45,1.45);
  \begin{scope}[shift={(7.1,0)}]
    \draw[axis] (0,0) -- (4.8,0) node[above left] {$\theta$};
    \draw[axis] (0,0) -- (0,3.4) node[below right, yshift=+5pt] {$p(\theta)$};
    \draw[dashedbox] (0.7,0) -- (0.7,2.9) -- (4.0,2.9) -- (4.0,0);
    \draw[curve, domain=0:4.5, samples=250, smooth]
      plot (\x,{2.7*exp(-(\x-2.35)^2/(2*0.55^2))});
  \end{scope}
  \end{tikzpicture}
  \caption{Imperfect approximation of a box prior leads to biased inference if NPE is used in an i.i.d.\ situation. Multiplication of NPE posteriors amplifies the approximation error.}
  \label{fig:npeiid}
\end{SCfigure}

It may seem daunting to choose among the collection of techniques that comprise SBI.
How to decide between NPE, NLE, NRE, how to set the hyperparameters, how to compress?

We only give a brief starting point and refer the reader to more in-depth discussions~\citep[e.g.,][]{Montel2025,Deistler2025}.
NPE should be the default choice thanks to its simplicity, speed, and ability to handle high-dimensional data vectors.
If inference is performed from multiple i.i.d.\ measurements, NPE should be avoided since it already contains the prior and thus in general does not allow for the factorization of the posterior
(note that the special case of uniform prior with sharp edges may seem like an exception, but due to the regularity properties of neural functions it would still introduce errors as illustrated in Fig.~\ref{fig:npeiid}).

Besides the i.i.d.\ situation, NLE may also be preferable in the less common situation of comparatively high-dimensional parameter vector.
NRE would be the natural choice in the special situations in which a flow architecture is inconvenient or too computationally expensive.

Compression is necessary in most realistic scenarios.
For tabular data such as the usual summary statistics, we recommend to begin with simple techniques such as MOPED and assess later whether neural compression is necessary.
For data with additional structure such as fields or images, we recommend to pay attention to the symmetry properties and inductive biases the compressor should possess.
Pretraining on similarly structured data may improve sample efficiency.

Hyperparameter tuning is an important part of SBI and should not be neglected.
This should proceed similarly to usual machine learning practices: loss curve interpretation, $k$-fold cross-validation.
Some of the diagnostics in the below Sec.~\ref{sec:diagnostics} can also be used in addition to the loss value.
However, it can be useful to hold out diagnostics for testing so as to not overfit on them.

\section{Diagnostics}\label{sec:diagnostics}

Any inference on model parameters should go through a series of diagnostic tests aimed at finding problems.
This is not unique to SBI, but the black-box character of neural networks implies that additional tests are warranted.
To understand these tests, it is useful to keep in mind two possible failure modes we can expect for the posterior.
The first failure mode is model misspecification, that is a systematic effect independent of the use of SBI.
The second failure mode is a poorly trained neural network even though the forward model is sufficiently accurate.
This can happen due to a small training set or suboptimal hyperparameter choices.
In many applications one can trade between these two failure modes (because simulation accuracy anti-correlates with speed).

\subsection{Basic tests}

\begin{SCfigure}
  \includegraphics[width=0.7\textwidth]{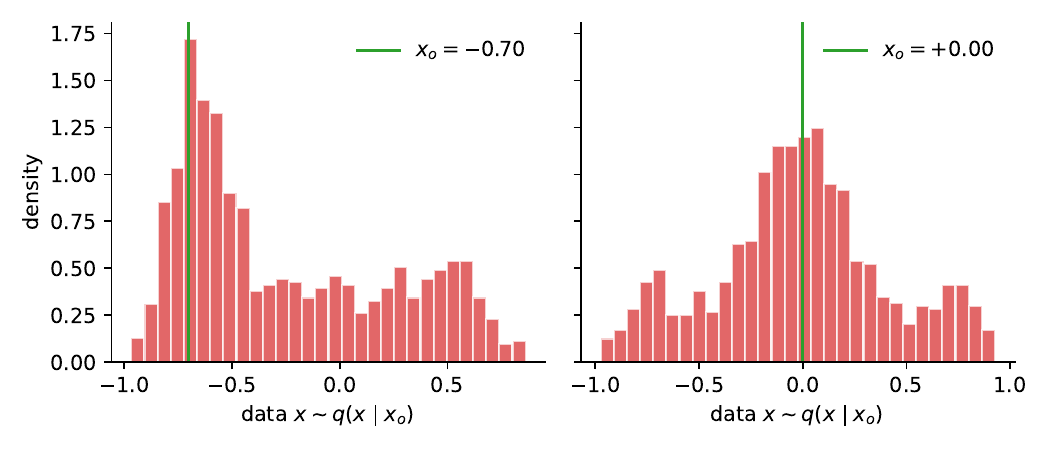}
  \caption{Example of posterior predictive checks. The observed $x_o$ should be ``typical'' of $q(x|x_o)$.}
  \label{fig:ppc}
\end{SCfigure}

Many tests carry over from explicit-likelihood analyses.
Data censoring can often uncover problems.
In cosmological inferences, this can for example mean removing some small-scale information or multipoles, restricting to certain patches on the sky, and similar tests.
Practitioners usually understand well how the posterior should qualitatively respond to such changes in the data vector.
If the neural posterior does not match the expectation, problems can be caught early.
A powerful extension on such tests is to compare the posterior's response in real data and mock data inferences.

Another basic (but computationally expensive) test are posterior predictive checks.
We draw new parameter vectors $\theta \sim p(\theta|x_o)$ from the neural posterior and run the forward model on them.
The resulting data vectors are then compared with the observed data.
For a correct posterior, the observed data should look like a ``typical'' sample from the ensemble.
In our one-dimensional example, this comparison is intuitive, see Fig.~\ref{fig:ppc}.
For higher-dimensional data, practitioners can come up with spot checks.

Both data censoring and posterior predictive checks are mixed tests in that they are sensitive to both failure modes described above.

\subsection{Calibration tests}\label{sec:coverage}

\begin{figure}
  \includegraphics[width=0.48\textwidth]{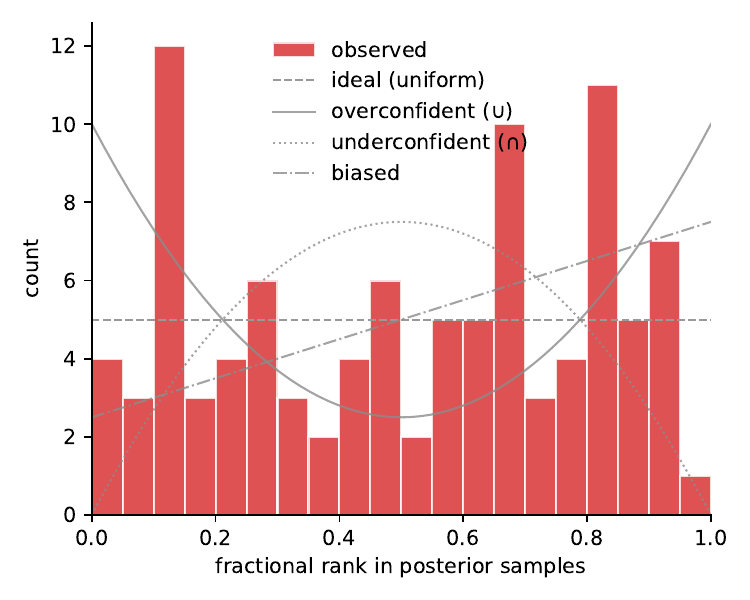}
  \includegraphics[width=0.48\textwidth]{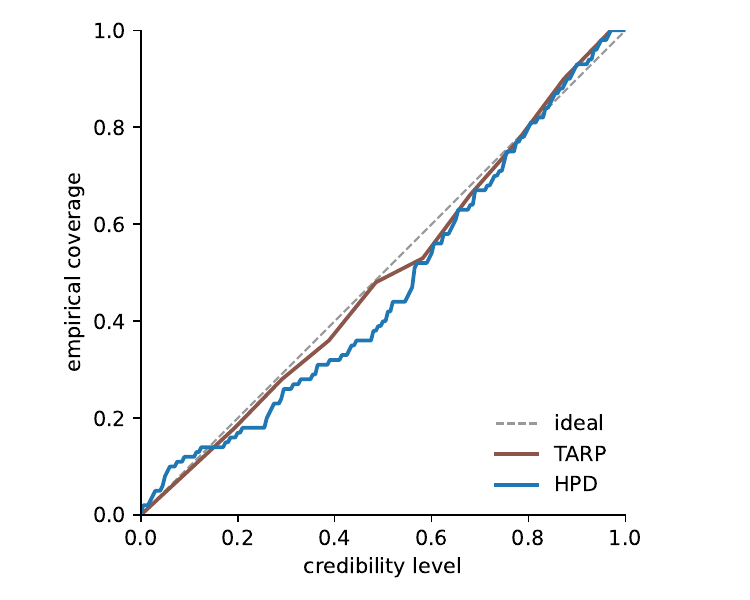}
  \caption{Illustration of calibration tests. \emph{Left:} ranks test, including illustrations of typical failure modes. \emph{Right:} coverage tests (highest posterior density, HPD, and distance to random point, TARP).}
  \label{fig:calibration}
\end{figure}

Calibration\index{calibration} tests isolate the second failure mode, namely undertraining of the neural network.
They are thus internal diagnostics without reference to the actual observed data and cannot probe for simulator misspecification.
All calibration tests rely on a test set of simulations drawn from the same joint $p(\theta, x)$ as the training set.
Note that this requirement on identical priors is the reason why sequential methods (Sec.~\ref{sec:sequential}) do not allow for direct application of the calibration tests described here.

All coverage\index{coverage} tests rely on some notion of ranking.
The idea behind this is rather simple.
Imagine we are given samples from the neural posterior, into which someone has mixed the true parameter vector.
Iff the neural network is trained perfectly and thus the posterior is well-calibrated, there exists no strategy that could identify the true parameter vector among all the samples with better than random chance.
It is easy to see that any attempted strategy is equivalent to ranking the samples in some way and picking the $N$th where $N$ is some pre-determined number.
Thus, we should test whether we can find a ranking scheme for which there is a statistically significant excess of true parameter vectors at any $N$th spot.
Since there are many ranking schemes, we concentrate on a few sensible ones.

The simplest calibration diagnostic is the ranks test~\cite{Talts2018} which is also known as ``simulation-based calibration'' (SBC).
For each simulation in the test set, inference is performed and the resulting set of samples is ordered according to one of the parameters.
Then we compute the rank of the true parameter value in this ordered chain.
For a well-calibrated posterior, this rank should be uniformly distributed over draws from $p(\theta, x)$.
We can test this hypothesis by computing the histogram of ranks over the test set, see the left panel in Fig.~\ref{fig:calibration}.
Systematically over- or under-confident posteriors manifest as characteristic $\cup$ and $\cap$ shapes.

A popular more complicated test is the coverage test using the highest posterior density (HPD) region~\cite{Hermans2022}.
In this test, the true parameter vector defines the boundary of a credible region within which the posterior density exceeds the value at the boundary (in the case of multimodal posteriors, there may be multiple disjoint parts to the credible region).
The integral under the thus defined credible region is a credibility level.
For a well-calibrated posterior, this credibility level should again follow a uniform distribution.
In this case, the ranking is according to posterior density.
HPD coverage can take into account multidimensional parameter vectors whereas the ranks test needs to be performed for each dimension separately.
Traditionally, the integrated histogram is plotted for HPD coverage, see the right panel of Fig.~\ref{fig:calibration}.

An alternative to the HPD coverage test is the test of accuracy with random points (TARP)~\cite{Lemos2023}.
Instead of defining the credible region by equal posterior density along its boundary, TARP constructs a credible region as a sphere whose center is a randomly sampled parameter vector and the radius is such that the true parameter vector lies on the surface.
This is again a ranking procedure according to the distance from a random point.
Note that TARP uses a metric on parameter space, thus a sensible normalization improves sensitivity.

So far, we have only discussed global calibration tests.
There can be cases in which the posteriors are mis-calibrated in such a way that the mis-calibration averages out when the entire distribution $p(\theta, x)$ is considered.
Such problems can be diagnosed with local coverage tests, such as local C2ST~\cite{Linhart2023}.
Of course, local tests require more test simulations and thus may not be possible in realistic cases.

\subsection{Posterior validation against ground truth}

There are cases in which neural posteriors can be validated against ground-truth posteriors.
The primary example is whenever SBI is used for performance reasons to replace an existing explicit-likelihood inference algorithm.
In such cases, we can directly validate the posteriors.
What is needed is a way to compare two sets of samples.

There are several approaches to this problem.
C2ST~\cite{lopezpaz2018} is a two-sample classifier which gets trained under supervision to distinguish between samples from neural and ground truth posteriors.
The posteriors are deemed identical if the classifier cannot surpass random-guessing performance substantially.
Maximum mean discrepancy (MMD)~\cite{gretton12a} maps the samples into a high-dimensional feature space (using the kernel trick) and compares the means of the two posteriors in this space.
In the high-dimensional embedding, means are expected to capture almost all moments of the lower-dimensional posteriors.
PQMASS~\cite{Lemos2025} is an alternative which performs well in moderately high dimensions but does not require choosing a specific kernel.

\section{Applications in Cosmology and Astrophysics}

The application of SBI to cosmology and astrophysics is a rapidly expanding area.
We thus only list a few representative papers.
For brevity, we only consider works in which SBI has been applied to real data.

In cosmology, currently the main application for SBI is in cases of intractable likelihoods, both at the field level and for summary statistics.
For galaxy redshift surveys, analyses have been performed only on BOSS data thus far~\citep[e.g.,][]{Hahn2024,Thiele2024}.
For weak gravitational lensing, analyses have been performed for all major Stage-III surveys:
DES~\cite{Jeffrey2021,Jeffrey2025,Gatti2025}, KiDS~\cite{Fluri2022,vonWietersheim-Kramsta2025}, and HSC~\cite{Novaes2025}.
Ref.~\cite{vonWietersheim-Kramsta2025} motivate their work partly through complicated systematic effects whose inclusion is difficult in explicit-likelihood techniques.

In astrophysics, the situation is more complex, in that both intractable likelihoods and performance considerations motivate the use of SBI.
Speed is the deciding factor in gravitational wave source inference~\cite{Dax2021}, supernova population level analyses~\citep[a rare hierarchical analysis,][]{Karchev2024}, photometric redshift inference~\cite{Wang2023}, and stellar parameter inference from spectroscopy~\cite{Zhang2023}.
On the other hand, an analysis of the galactic center $\gamma$-ray excess is motivated by an intractable likelihood~\cite{Mishra-Sharma2022}.

\section{Current frontier}

There are broadly speaking two directions in which SBI research is currently advancing.

The first direction concerns model expressivity and inductive biases.
In some applications, the limited expressivity of the discrete-time normalizing flows that are commonly used in NPE represents an obstacle.
Thus, alternatives are being developed which may suffer from slow density evaluations and thus require further research.
In cosmology and astrophysics, typical posteriors are relatively simple (although there are exceptions).
Thus, this first research direction may be considered less relevant.

The second direction concerns posterior fidelity in the realistic scenario of finite training sets.
As noted in several works~\cite{Hermans2022,Homer2024,Bairagi2025}, with limited simulation budgets the inferred posteriors can easily be quite wrong.
Since cosmological and astrophysical simulations are typically extremely expensive, research on improving the performance in this regime is very relevant.

One approach to ensure faithful inference with finite training sets is to improve coverage (see Sec.~\ref{sec:coverage}).
This can be done via ensembling of multiple networks~\cite{Alsing2019,Hermans2022}, through a learning objective that directly targets coverage~\cite{Falkiewicz2023}, or via balancing~\cite{Delaunoy2022,Delaunoy2023}.

A different approach to inference in the simulation-limited regime is to augment the small training set with other information.
One possibility is to identify parts of the likelihood which can be well approximated through analytic means and thus isolate a simpler SBI problem.
In cosmology, this is quite natural since large-scale evolution is close to linear and thus well-understood~\cite{modi2023hybridsbiilearned,Zhang2025}.
Another possibility is to augment the training set with a large number of cheaper approximate simulations and thus perform multi-fidelity training~\cite{Krouglova2025,Thiele2025,Hikida2025}.
Finally, one can rely entirely on the cheaper simulations for training and use a small number of accurate simulations for a correction step~\cite{Wehenkel2024,Jia2024b,Ruhlmann2025}.

\section{Summary and Outlook}

The use of SBI methods is motivated in two types of situations, namely when facing intractable likelihoods and when sampling from the posterior is time-critical.
At the moment, in cosmology the first situation is more common, while the second situation is more common in astrophysics.
We have introduced the main techniques (NPE, NLE, NRE) and discussed some alternatives.
A critical part of the SBI workflow is running as many diagnostics as possible, the more common ones of which we have described.

Obtaining accurate posteriors from limited simulation budgets is the most important problem to be solved going forward.
We have mentioned some current directions the field is pursuing towards this goal.
If robust limited-budget training can indeed be realized, SBI will gain the community's trust as an inference tool on par with traditional explicit-likelihood sampling.

Cosmology is a challenging field to apply SBI in.
Simulations are expensive and there is only one realization of the data-generating process (namely our universe).
At the same time, there are some analyses we so far have only identified SBI as a plausible technique for.
Thus, we expect SBI to continue playing an important role in the future of cosmological inference.
Astrophysics appears a more natural field for SBI since in many cases there are catalogs of approximately i.i.d.\ data vectors.
On the other hand, the physics tends to be more complex than on large cosmological scales and thus simulators can be difficult to construct.
It seems likely that SBI will play an important role in some areas such as gravitational waves while advancing more slowly in areas where simulations are still being refined.

\begin{acknowledgement}
I would like to thank my collaborators at CPPM for their warm hospitality while this chapter was being finished.
LT is supported by JSPS under KAKENHI 24K22878 and 26K17136 and by the Royal Society under ICA\textbackslash R2\textbackslash 252140.
\end{acknowledgement}

\bibliography{chapter}


\end{document}